\documentclass[aps,prl,twocolumn,superscriptaddress,showpacs]{revtex4}
\usepackage{graphicx}
\usepackage{mathrsfs}
\usepackage{amssymb,amsmath}
\usepackage{bm}

\begin{document}

\title{Two-Dimensional Topological Insulator State and Topological
  Phase Transition 
in Bilayer Graphene}

\author{Zhenhua Qiao $^*$}
\affiliation{Department of Physics, The University of Texas at
Austin, Austin, Texas 78712, USA}
\author{Wang-Kong Tse $^*$}
\affiliation{Department of Physics, The University of Texas at
Austin, Austin, Texas 78712, USA}
\author{Hua Jiang}
\affiliation{International Center for Quantum Materials, Peking
University, Beijing 100871, China} \affiliation{Department of
Physics, The University of Texas at Austin, Austin, Texas 78712,
USA}
\author{Yugui Yao}
\affiliation{Institute of Physics, Chinese Academy of Sciences,
Beijing 100190, China} \affiliation{Department of Physics, The
University of Texas at Austin, Austin, Texas 78712, USA}
\author{Qian Niu} \affiliation{Department of
Physics, The University of Texas at Austin, Austin, Texas 78712,
USA} \affiliation{International Center for Quantum Materials, Peking
University, Beijing 100871, China} 
\begin{abstract}
We show that gated bilayer graphene hosts a strong topological 
insulator (TI) phase in the presence of Rashba spin-orbit (SO) coupling. 
We find that gated bilayer graphene under preserved time-reversal symmetry is a
quantum valley Hall insulator for small Rashba SO coupling $\lambda_{\mathrm{R}}$,
and transitions to a strong TI when $\lambda_{\mathrm{R}} > \sqrt{U^2+t_\bot^2}$,
where $U$ and $t_\bot$ are respectively the interlayer potential 
and tunneling
energy. Different from a conventional quantum spin Hall state, the edge modes
of our strong TI phase exhibit both spin and valley filtering, and
thus share the properties of both quantum spin Hall and quantum valley
Hall insulators. 
The strong TI phase remains robust in the presence of weak graphene intrinsic SO coupling. 
\end{abstract}

\maketitle

Recently, there has been a surge of interest in time-reversal
invariant topological insulators (TI) \cite{TI_Rev}, a new quantum phase of
matter that carries an odd number of helical edge (two-dimensional
TIs) or surface (three-dimensional TIs) states. Two-dimensional TI, 
commonly known as quantum spin Hall (QSH) insulator, occurs in strongly
spin-orbit coupled material and was predicted in single-layer graphene with intrinsic spin-orbit (SO)
coupling \cite{Kane} and in HgTe/CdTe quantum well at large well thicknesses 
\cite{QSH_Zhang}. The latter has been confirmed in experiment
\cite{QSH_Zhang_exp}; graphene, however, has a weak intrinsic SO
coupling \cite{WeakSOI}, making it difficult to observe a QSH state. 
To remedy the situation, a number of recent theoretical
\cite{Guinea_SO, qiao, James, Franz, HBZhang} and experimental \cite{RashbaSOI}
work have demonstrated that surface doping on graphene with heavy
atoms can dramatically boost the SO coupling strength. Moreover, the
broken out-of-plane mirror symmetry creates strong Rashba SO coupling
\cite{RashbaSOI}, which can induce an interesting quantum anomalous 
Hall state \cite{qiao,James} in the presence of proximity magnetic exchange 
interaction. 

In this Letter, we present a theory of topological phases in gated
bilayer graphene in the presence of Rashba SO coupling $t_{\mathrm{R}}$ under preserved 
time-reversal symmetry. From arguments of band structure and $\mathbb{Z}_2$ topological invariant, we show that this gated bilayer system exhibits two
topologically distinct phases, from a quantum valley Hall state at
weak $t_{\mathrm{R}}$ to a strong topological insulator state at
strong $t_{\mathrm{R}}$. In a zigzag-edged bilayer system, the strong 
TI phase has the properties of both quantum valley Hall and quantum spin Hall states. 
At a fixed $t_{\mathrm{R}}$, topological phase 
transition between the two states can be achieved by gate tuning. We
also show that the strong TI phase remains robust if weak intrinsic SO coupling is present in addition to the
Rashba effect.  

\begin{figure*}
\includegraphics[width=17cm,totalheight=11cm,angle=0]{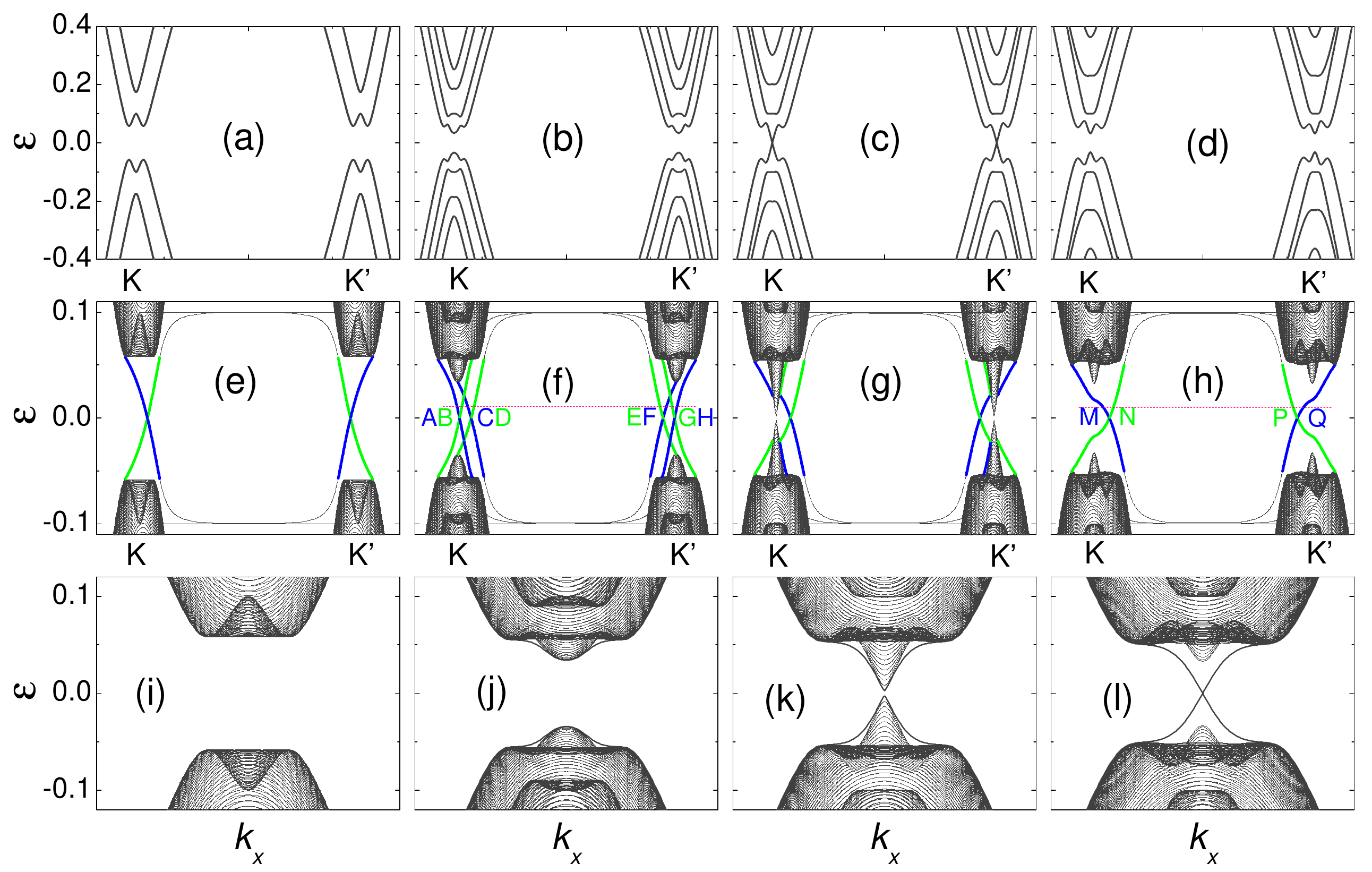}
\caption{Evolution of band structure of gated bilayer graphene at a 
  fixed interlayer bias $U/t=0.1$ for increasing Rashba SO coupling
  $t_{\mathrm{R}}/t = 0, 0.04, 0.0582, 0.08$. $t_{\mathrm{R}}$ is
  assumed to be the same on both layers for concreteness. First row (panels a-d):
  bulk system with periodic boundary conditions; second row (panels
  e-h): finite strip with zigzag edges; third row (panels i-l): finite
  strip with armchair edges. In the second row, the dark/blue and
  light/green curves inside the bulk gap are used to represent edge
  states located at opposite boundaries. $k_x$ is normalized to
  inverse lattice constant $1/a$ and the valleys are indicated as K, K'.
} \label{Fig1}
\end{figure*}

The tight-binding Hamiltonian for the AB-stacked bilayer 
graphene \cite{BLG} in the presence of Rashba SO coupling and interlayer
potential difference (due to an applied gate voltage) is \cite{James,Smith}
\begin{eqnarray}
H_{\rm BLG} &=& H_{\rm SLG}^{\mathrm{T}}+H_{\rm SLG}^{\mathrm{B}}+t_\bot \sum_{i\in \mathrm{T}, j \in \mathrm{B}, \alpha}{
  c_{i\alpha}^\dagger c_{j
\alpha}} \nonumber \\
&&+U\sum_{i\in \mathrm{T},\alpha}{ c_{i\alpha}^\dagger c_{i
\alpha}}-U\sum_{i\in \mathrm{B},\alpha}{ c_{i\alpha}^\dagger c_{i
\alpha}}, \label{BilayerH}
\end{eqnarray}
where the single-layer Hamiltonian $H_{\rm SLG}^{\mathrm{T,B}}$ for the top (T) and bottom (B) graphene
layers including Rashba SO coupling \cite{Kane,Wrinkler} is 
\begin{equation}
H_{\rm SLG} =t\sum_{\langle{ij}\rangle \alpha }{ c^\dagger_{i
\alpha}c_{j\alpha}+ {i}t_{\mathrm{R}}\sum_{\langle{ij}\rangle \alpha \beta
}(\bm{s}_{\alpha \beta}{\times}\bm{d}_{ij})_z
c_{i \alpha }^\dagger c_{j \beta }},\label{singleH}
\end{equation}
where $c_{i\alpha}^\dagger$ is the usual creation operator for electron with
spin $\alpha = \pm 1$ on site $i$ and $t$ is the intralayer tunneling
energy between nearest neighhor sites. The second term on the right-hand
side is the Rashba SO interaction with coupling strength
$t_{\mathrm{R}}$, $\bm s$ is the Pauli matrices for the spin degrees
of freedom, and $\bm{d}_{ij}$ is the lattice vector pointing from site
$j$ to site $i$. Interlayer tunneling between the two layers is given
by the third term in Eq.~(\ref{BilayerH}) with a tunneling energy 
$t_\bot$, whereas interlayer potential difference $2U$ is given by the last
two terms. 

We first analyse the bulk band structure obtained from the above Hamiltonian.
Figs.~\ref{Fig1}a-d shows the evolution of the bulk band
structure with increasing strength of Rashba SO coupling at fixed
interlayer potential difference. 
In bilayer graphene, a bulk band gap can be opened (panel a) by applying an
external gate voltage across the layers 
\cite{BLG_bandgap_theory_exp}   
to break the inversion symmetry in the out-of-plane direction. When the Fermi level lies within the bulk gap, gated
bilayer graphene is a quantum valley Hall (QVH) insulator
\cite{James,QVH_Ref}, 
characterized by a quantized valley Chern number $C_{\mathrm{v}}$,
which is defined as the difference between the Chern numbers at the two
valleys K and K'. 
When the Rashba SO coupling $t_{\mathrm{R}}$ is turned on, we find that the bulk
gap decreases gradually with $t_{\mathrm{R}}$ (panel b) and 
vanishes completely (panel c). Since turning on the Rashba
coupling from zero is not accompanied by any band gap closing, it can
be inferred that the system remains a QVH insulator at finite
$t_{\mathrm{R}}$ and $U$, before the bulk gap vanishes in panel c. 
We find that the bulk gap reopens (panel d) when $t_{\mathrm{R}}$ is
further increased, and in the vicinity of gap closing 
the conduction and valance bands cross each other linearly as a function of $t_{\mathrm{R}}$
characteristic of a band inversion. This suggests a topological phase transition, and in the
following we show that is indeed so with the emergent phase 
a two-dimensional strong topological insulator that, interestingly, also possesses 
the properties of a QVH insulator in the sense that the $\mathbb{Z}_2$ invariant \cite{FuKane_Z2} is $1$ 
and the valley Chern number $C_{\mathrm{v}}$ is also $1$. 

\begin{figure}
\includegraphics
[width=8.5cm,angle=0]{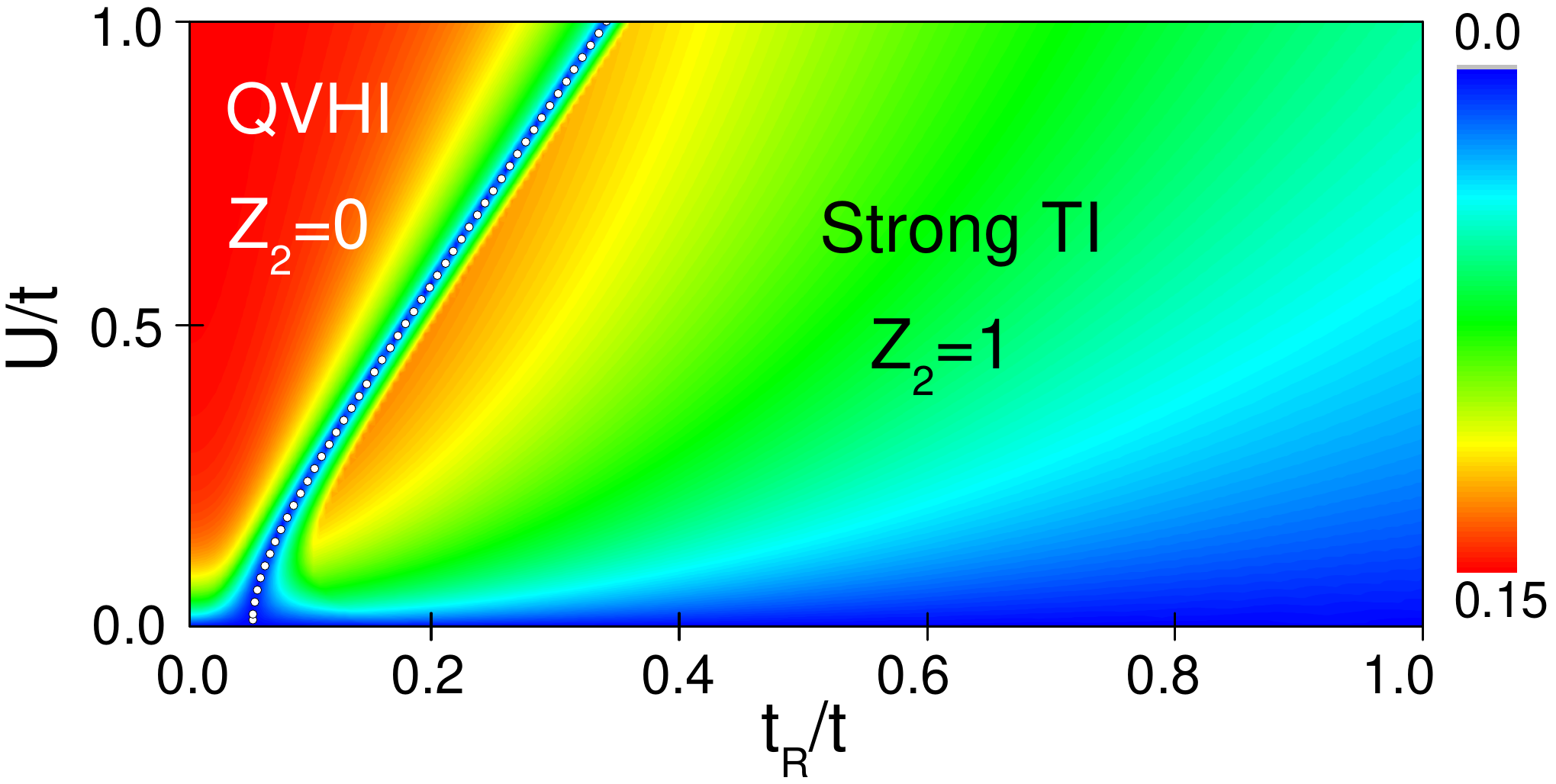}
\caption{(Color online) Phase diagram of the $\mathbb{Z}_2$ invariant
  as a function of $U$ and $t_{\mathrm{R}}$ at fixed interlayer tunneling $t_\bot/t=0.1429$. The color scale
represents the magnitude of the bulk gap in units of $t$. The dotted line
plots the phase boundary condition Eq.~(\ref{relation}) between the
quantum valley Hall insulator (QVHI) and strong TI 
phases. The bulk gap decreases but remains finite at $U \neq 0$ as $t_{\mathrm{R}}$
increases to large values (dark/blue region below the light/green
region) in the strong TI phase. 
} \label{phase}
\end{figure}

The $\mathbb{Z}_2$ invariant \cite{FuKane_Z2} characterizes the band topology in the
presence of time-reversal symmetry and is defined by 
\begin{equation}
\mathbb{Z}_2 = \frac{1}{2\pi}\left[\oint_{\partial\,
    \mathrm{HBZ}}\mathrm{d}\bm{k}\cdot\bm{A}(\bm{k})-\int_{\mathrm{HBZ}}\mathrm{d}^2k\,\Omega_z(\bm{k})\right]\,\mathrm{mod}(2),
\label{Z2}
\end{equation}
where $\bm{A}(\bm{k}) = i\sum_n\langle u_n(\bm{k})\vert\nabla_k u_n(\bm{k})\rangle$ is the
Berry connection summed over all filled band indices $n$ with the periodic part
of the Bloch function denoted by $\vert u_n(\bm{k})\rangle$, $\Omega_z(\bm{k}) =
(\nabla_k\times \bm{A})_z$ is the $z$ component of the Berry
curvature. By virtue of Kramer's theorem $\vert u_n(\bm{k})\rangle$
satisfies the time-reversal invariant constraint $\vert u_n(-\bm{k})\rangle = \Theta\vert
u_n(\bm{k})\rangle$, where $\Theta$ is the time-reversal operator.
Therefore, we only need to calculate the line and surface integrals in
Eq.~(\ref{Z2}) over half of the Brillouin zone (as denoted by `HBZ' in the equation) that satisfies the time-reversal constraint. We have computed $\mathbb{Z}_2$
numerically from the Hamiltonian Eq.~(\ref{BilayerH}) using the method described in Ref.~\cite{Z2_method}. Fig.~\ref{phase} shows our
calculated $\mathbb{Z}_2$ phase diagram as a function of $U$ and
$t_{\mathrm{R}}$, where 
we find that the regimes before and after gap closing are
characterized by topologically distinct phases. The system before gap
closing is in a QVH phase with a topologically trivial 
$\mathbb{Z}_2 = 0$ invariant. After gap closing and reopening, we find
that $\mathbb{Z}_2 = 1$, therefore proving that the gated bilayer
graphene system is a strong TI. 

At the topological phase transition critical point, the gap closing
condition 
allows us to obtain an analytic expression of the phase transition boundary
from a low-energy Hamiltonian. 
Expanding the tight-binding Hamiltonian Eq.~(\ref{BilayerH}) in the
vicinity of K, K' 
gives the following eight-band low-energy Hamiltonian
\begin{eqnarray}
H&=& v (\eta \sigma_x k_x + \sigma_y k_y )\bm{1}_s\bm{1}_\tau +
\frac{t_\bot}{2}  ( \sigma_x \tau_x- \sigma_y \tau_y) \bm{1}_s 
\nonumber \\ &+&\frac{\lambda_{\mathrm{R}}}{2}(\eta\sigma_x s_y - \sigma_y s_x)
\bm{1}_\tau+U \bm{1}_\sigma \bm{1}_s \tau_z, \label{EffH}
\end{eqnarray}
where $\eta =\pm1$ labels the valley K, K' degrees of
freedom, $\bm{\sigma},~\bm{s}$ and $\bm{\tau}$ are Pauli matrices
representing the A-B sublattice, 
spin, and layer degrees of freedom, respectively; $\bm{1}$ is the
identity matrix, the Fermi velocity and Rashba coupling are given
respectively by 
$v = 3ta/2$ and $\lambda_{\mathrm{R}} = 3t_{\mathrm{R}}$. The low-energy Hamiltonian at $k=0$ gives the energy
eigenvalues $\varepsilon=\pm U$ and 
six other eigenenergies that satisfy the relationship
${\varepsilon}^3-\mu U({\varepsilon}^2  +\alpha^2-t^2_\bot-U^2)-
(\alpha^2+t^2_\bot+U^2)\varepsilon=0$
where $\mu$=$\pm 1$. Imposing the gap closing condition
$\varepsilon$=0 we find the topological phase transition boundary
\begin{equation}
\lambda_{\mathrm{R}}^2=U^2+t_{\bot}^2. 
\label{relation}
\end{equation}
In Fig.~\ref{phase} we plot Eq.~(\ref{relation}) on the $\mathbb{Z}_2$ phase
diagram, from which we see that the analytic expression (dotted line)
describes accurately the phase transition boundary between  
the two phases obtained from our numerical $\mathbb{Z}_2$ calculations.

\begin{figure}
\includegraphics
[width=8.5cm,totalheight=4cm,angle=0]{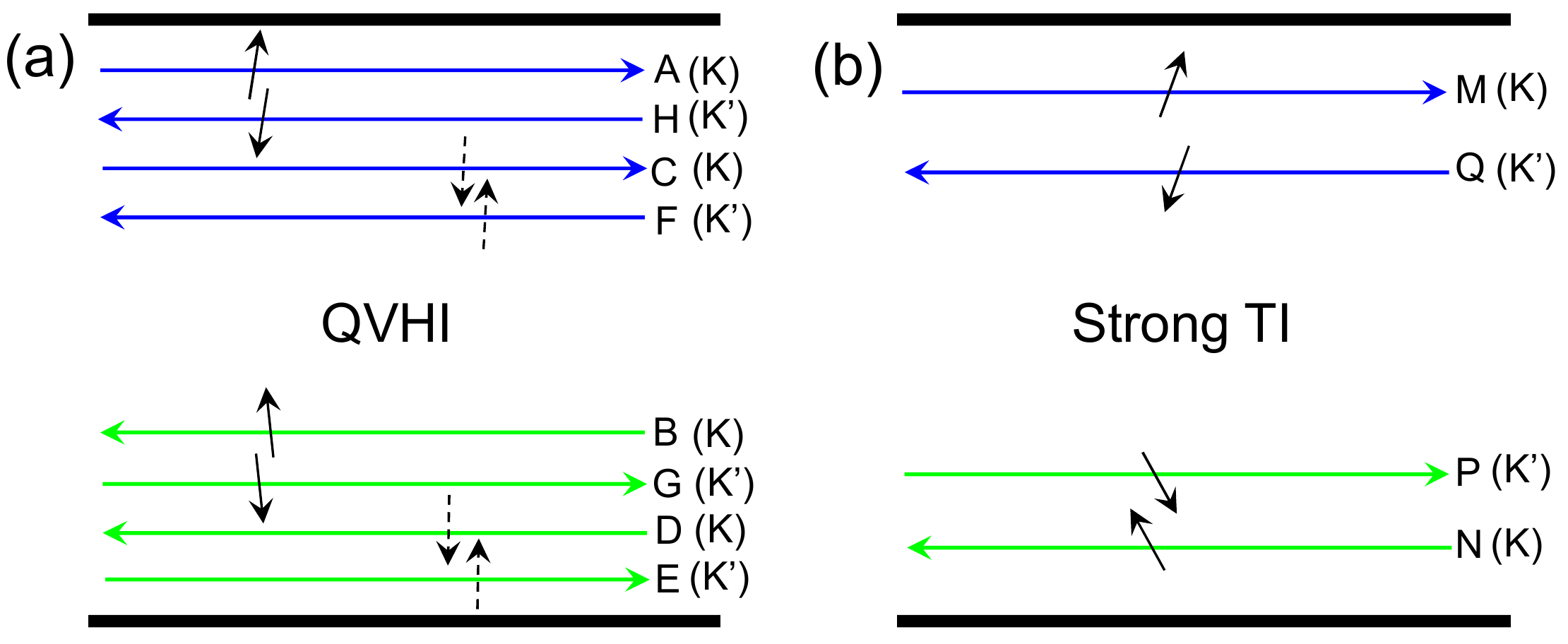}
\caption{(Color online) Schematic of edge state propagation in the 
  zigzag edge geometry for (a) QVHI phase at small $t_{\mathrm{R}}$;
  (b)  strong TI state at large $t_{\mathrm{R}}$. The arrows on 
  the edge channels represent in-plane spin directions (out-of-plane
  spin component is zero). 
Labels A-H and M-Q correspond to band
  labels in Figs.~\ref{Fig1} f and h.
}
\label{EdgeState}
\end{figure}
Graphene sheets have two principal edge terminations along and
perpendicular to the bond-length direction, respectively known as armchair and zigzag terminations \cite{GrapheneRev}. The valleys K, K' remain good quantum numbers in
zigzag-edged strips but are mixed (and hence no longer good
quantum numbers) in armchair-edged strips. 
We first examine the edge band structure in a bilayer graphene 
strip 
with zigzag edges along
one direction and periodic boundary condition along the other
direction.
Fig.~\ref{Fig1}e shows the QVH phase at finite $U$ and $t_{\mathrm{R}}
= 0$ characterized 
by a pair of spin-degenerate gapless edge bands. 
We find that the two valleys are characterized by opposite Chern
numbers $\pm 1$, therefore the valley Chern number
$\mathcal{C}_{\mathrm{v}} = 2$. 
With a finite Rashba SO coupling (panel f), the spin degeneracy is 
lifted yielding two separate pairs of gapless edge bands, and 
the bilayer system remains a QVH insulator with the same valley Chern
number $\mathcal{C}_{\mathrm{v}} = 2$. 
It can be seen that the outer pair of edge bands (e.g.  at valley
K, lines labeled A, B) connect the conduction band with the valance band at the same valley, whereas the 
inner pair of edge bands (e.g. C, D at valley K) connect the two
conduction bands or the two valence bands at different valleys. 
When the bulk gap is closed (panel g), 
the two pairs of
edge bands at each valley 
merge together with the bulk bands (the upward and downward dips at K
and K'); when the bulk gap reopens (panel h) at a larger
$t_{\mathrm{R}}$, only one pair of non-degenerate edge states emerges. 
This change from an even to an odd number of edge states signals a
phase transition from a topologically trivial to a
topologically 
nontrivial phase, consistent with our $\mathbb{Z}_2$
calculation. Remarkably, we find that the valley Chern number remains
quantized, but changes to $\mathcal{C}_{\mathrm{v}} =
1$. This implies that the strong TI phase is also
a QVH insulator and enjoys the same valley protection. 

This is illustrated in Fig.~\ref{EdgeState} showing the edge modes of the QVH and strong TI
phases before and after bulk gap closing. At small $t_{\mathrm{R}}$,
the QVH phase (Fig.~\ref{EdgeState}a) 
has two pairs of counterpropagating edge states on each edge that are
valley-filtered with different valley quantum numbers K and K'. At 
large $t_{\mathrm{R}}$ after gap reopening the strong TI phase
carries only a single pair of counterpropagating edge
states. Although the $z$ projections of spins are not good quantum 
numbers, these counterpropagating edge channels still carry helically opposite spins that are rotated from $s_z$ due to Rashba SO coupling. 
In the conventional QSH 
phase \cite{Kane}, the counterpropagating edge states constitutes a
Kramer's pair that are spin-filtered. 
A novel feature in our strong TI phase is that because of valley quantum number conservation, the
pair of counterpropagating edge states are \textit{both} spin-filtered
and valley-filtered (Fig.~\ref{EdgeState}b), consistent with our bulk
topological invariant results $\mathbb{Z}_2 = 1$ and $C_{\mathrm{v}} = 1$.
As a consequence, the strong TI phase is topologically protected both by
time-reversal symmetry against weak non-magnetic disorder, and by 
 valley-inversion symmetry against weak magnetic disorder that is
 long-range (longer than lattice spacings) so that intervalley
 scattering remains prohibited. 

For armchair edge geometry, because valleys K and K' overlap and are
not good quantum numbers, there is no QVH phase. At finite $U$ and
small $t_{\mathrm{R}}$ before phase transition, the system is an
ordinary insulator and does not have any gapless edge state (Fig.~\ref{Fig1} i - j).
When the bulk gap closes and reopens (panels k -l), a single pair of gapless edge states 
emerges that are not valley-filtered but remain spin-filtered, as
expected from a strong TI phase. 
Unlike the zigzag case however, the armchair case has no 
valley protection and thus carries a strong TI phase akin to the conventional QSH state.  

\begin{figure}
\includegraphics
[width=8.5cm,angle=0]{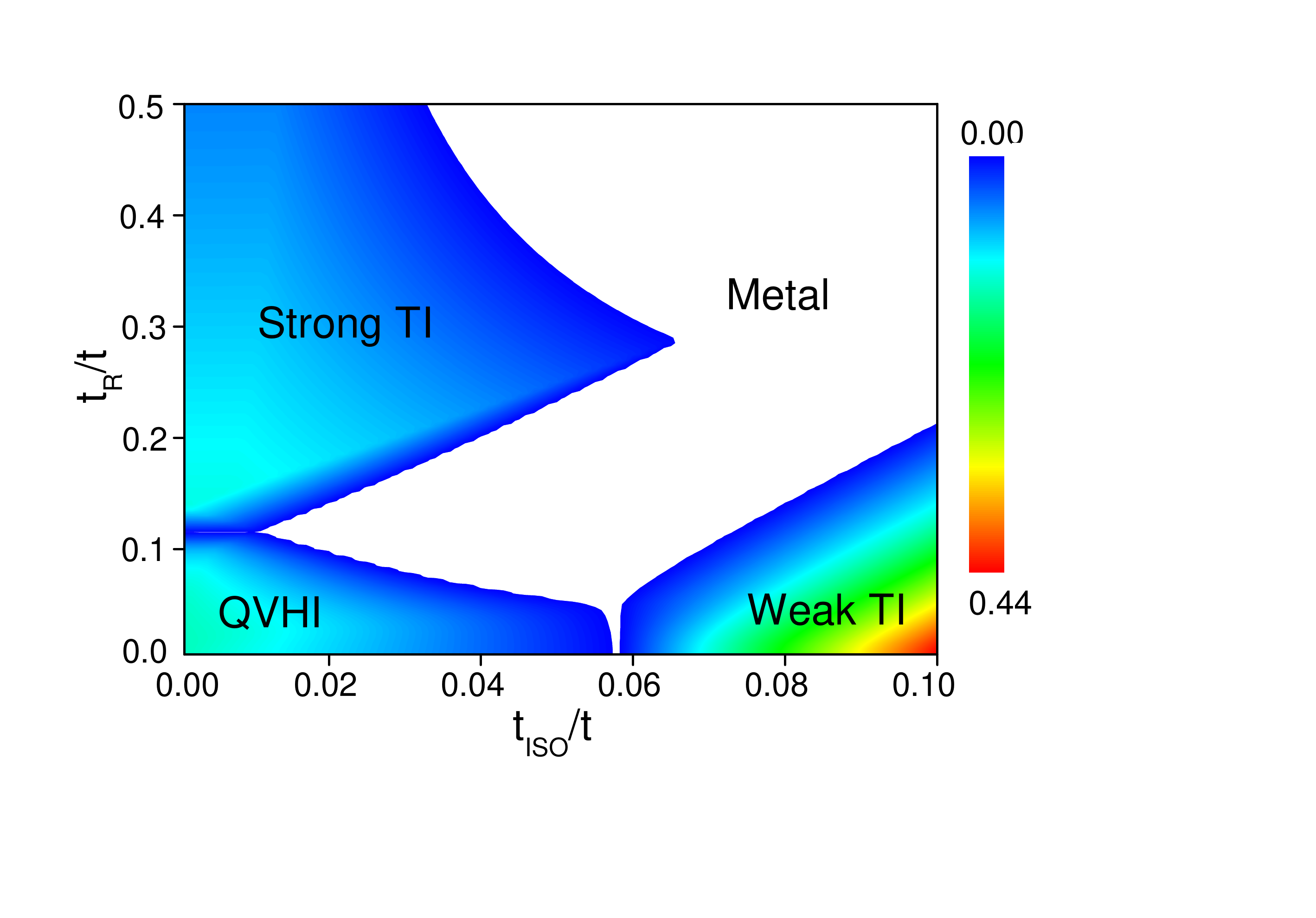}
\caption{(Color online) Phase diagram as a function of Rashba
  $t_{\mathrm{R}}$ and intrinsic $t_{\mathrm{ISO}}$ SO coupling
  strengths (taken to be same on both layers) at finite $U/t =
  0.3$. The color scale indicates the 
  magnitude of the bulk gap in units of $t$. The white region
  corresponds to a metallic phase where there is no global gap in the
  bulk band structure.
} \label{Fig3}
\end{figure}

The predicted TI state in this Letter relies on the presence of a strong 
Rashba SO coupling, which can be achieved in principle through 
doping with adatoms \cite{RashbaSOI, qiao, James, Franz}. 
This however also enhances the intrinsic SO coupling, and
therefore leads to a natural question whether or not the TI state will
be destroyed by the presence of intrinsic SO coupling.
We address this question by including 
the intrinsic SO coupling term \cite{ISO_Ham} in each layer of the Hamiltonian Eq.~(\ref{BilayerH}). 
Fig.~\ref{Fig3} shows the phase diagram we obtained 
as a function of both Rashba SO and intrinsic SO 
coupling strengths at a fixed interlayer potential. First, for small 
$t_{\mathrm{R}}$, we find that the QVH phase remains intact when the
intrinsic SO coupling $t_{\mathrm{ISO}}$ is also small. As $t_{\mathrm{ISO}}$ in each layer is
increased, the individual-layer quantum spin Hall state due to 
intrinsic SO coupling prevails, leading to a phase transition to a weak TI phase 
\cite{Fertig_QSH} which is analogous to a layered QSH system. Despite
each layer behaves as a QSH state, an even number of such layers renders the overall system topologically trivial that is 
characterized by a vanishing $\mathbb{Z}_2$ and an even number of
gapless edge states. For large values of $t_{\mathrm{R}}$, we identify
a region in the phase diagram where the strong TI phase remains
robust. This occurs when the intrinsic SO coupling is about an order
of magnitude weaker than the Rashba SO coupling. Indeed, at small values 
of $t_{\mathrm{ISO}}$ the phase diagram remains qualitatively similar
to Fig.~\ref{phase} at $t_{\mathrm{ISO}} = 0$, with the only difference
that the gapless metallic regime (white region in Fig.~\ref{Fig3})
between the QVH and strong TI phases becomes more extended. 

In conclusion, we have shown that gated bilayer graphene hosts a
strong topological insulator phase at large Rashba spin-orbit
coupling. The gate voltage can serve as a topological switch that
tunes between the quantum valley Hall phase and the strong topological
insulator phase. This can be realized by enhancing the spin-orbit
coupling in graphene through adatom doping.

\emph{Acknowledgements ---} We acknowledge financial support from NSF 
(DMR 0906025), DOE (DE-FG03-02ER45958, Division of Materials Science and
Engineering), NSI-SWAN, Welch Foundation (F-1255, F-1473), and Texas Advanced Research Program. Y.Y. was supported by NSF of China (Grants
No. 10974231,11174337) and the MOST Project of China (Grants
No.  2011CBA00100).

$^*$ These authors contributed equally to this work.


\begin{thebibliography}{99}
\bibitem{ISO_Ham} The graphene intrinsic SO coupling is given by
  Eq.~(6) of Ref.~\onlinecite{Kane} with the same notations as our paper.
\end{thebibliography}
\end{document}